# The pre-log of Gaussian broadcast with feedback can be two

Michèle A. Wigger
Signal and Information Processing Laboratory
ETH Zürich
CH-8092 Zürich, Switzerland
Email: wigger@isi.ee.ethz.ch

Michael Gastpar
WiFo, Department of EECS
University of California, Berkeley,
Berkeley, CA 94720-1770, USA
Email: gastpar@berkeley.edu

*Abstract*—A generic intuition says that the pre-log, or multiplexing gain, cannot be larger than the minimum of the number of transmit and receive dimensions. This suggests that for the scalar broadcast channel, the pre-log cannot exceed one. By contrast, in this note, we show that when the noises are anti-correlated and feedback is present, then a pre-log of two can be attained. In other words, in this special case, in the limit of high SNR, the scalar Gaussian broadcast channel turns into two parallel AWGN channels. Achievability is established via a coding strategy due to Schalkwijk, Kailath, and Ozarow.

## I. INTRODUCTION

The significance of feedback in a capacity sense has been thoroughly studied for point-to-point and several network scenarios. Many results point to the lack of such a significance, starting with Shannon's proof that the capacity of a memoryless channel is unchanged by feedback. For networks, even for memoryless ones, feedback can increase capacity, as first shown by Gaarder and Wolf [1]. However, in most cases, the increase in capacity due to feedback remains modest, as expressed for example in a general conjecture in [2].

The exact feedback capacity remains unknown for most networks, with the notable exception of the two-user Gaussian multiple-access channel (MAC), whose capacity was found by Ozarow [3]. Some recent progress concerns the $M$-user Gaussian MAC [4]. Again, these results emphasize the lack of significance of feedback in a capacity sense.

By contrast, the result presented in this short note shows that feedback *can* have a rather significant impact on capacity in a certain *broadcast* setting. More specifically, we consider the problem of two-user broadcast subject to additive white Gaussian noise. This scenario has been studied previously by Ozarow [5], Ozarow and Leung [6], as well as Willems and van der Meulen [7].

The main result of this paper is that for the special case where the two noises are (fully) *anti-correlated*, in the limit as the available power becomes large, the trade-off between the two broadcast clients vanishes, and each client attains a rate as if the other did not exist. Formally, the result is presented as a "pre-log," or "multiplexing gain."

To our knowledge the considered setting is the first example of a channel where the "pre-log" is larger than the number of transmit antennas. This behavior is surprising in view of the result by Telatar [8] who showed that for *uncorrelated* noise sequences, the "pre-log" is upper bounded by the number of transmit antennas and by the number of receive antennas, even if the two receivers are allowed cooperate. Therefore, in the setting at hand when the noise sequences are uncorrelated the "pre-log" is upper bounded by one.

An extension of our result concerns the two-user Gaussian interference channel where all channel gains are equal. Again, we can show that when feedback is available and the noises are (fully) anti-correlated, the "pre-log" is two. Without feedback the "pre-log" is one, irrespective of the noise correlation.

The only situation known to date where a "pre-log" of two is achievable for a two-user Gaussian interference channel (only for non-equal channel gains) is when both transmitters know the other transmitter's message, which corresponds to a setting where the two transmitters can fully cooperate. If only one of the two transmitters knows the other transmitter's message the "pre-log" remains one [9]. Hence, this specific form of limited transmitter-cooperation does not increase the "pre-log". Our result here shows that in general limited transmitter-cooperation can be sufficient to increase the "pre-log" to two, e.g., when the limited transmitter-cooperation is established through full causal feedback links. For interference networks with more than two users the fact that limited transmitter-cooperation can increase the "pre-log" has been observed in [10] for the case where some of the transmitters know some of the other transmitters' messages.

One motivation for the study of anti-correlated noises is that the signals $Z_1$ and $Z_2$ in Figure 1 are due to one and the same outside interferer, but appear with different (more precisely, opposite) phase shifts at the two receivers.

## II. THE MODEL

The communication system studied in this note is illustrated in Figure 1. For a given time-$t$ channel input $x_t$ the channel outputs observed at receivers 1 and 2 are

$$Y_{k,t} = x_t + Z_{k,t}, \quad k \in \{1,2\}, \qquad (1)$$

where the sequence of pairs of random variables $\{(Z_{1,t}, Z_{2,t})\}$ is drawn independently and identically distributed (iid) for a

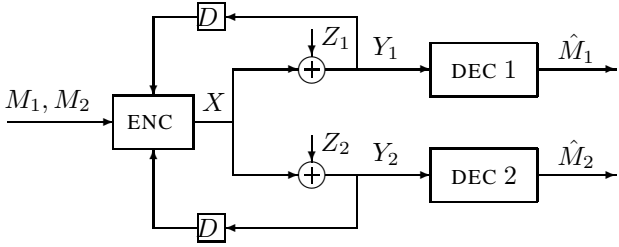

Fig. 1. The two-user AWGN broadcast channel with full causal output feedback.

normal distribution with zero mean and covariance matrix

$$\mathsf{K} = \begin{pmatrix} \sigma_1^2 & \rho_z \sigma_1 \sigma_2 \\ \rho_z \sigma_1 \sigma_2 & \sigma_2^2 \end{pmatrix} \quad (2)$$

for $\sigma_1, \sigma_2 > 0$ and $-1 \leq \rho_z \leq 1$.

The goal of the transmission is to convey message $M_1$ to Receiver 1 and an independent message $M_2$ to Receiver 2, where $M_1$ is uniformly distributed over the set $\{1, \ldots, \lfloor 2^{nR_1} \rfloor\}$ and $M_2$ is uniformly distributed over the set $\{1, \ldots, \lfloor 2^{nR_2} \rfloor\}$, $n$ being the block-length and $R_1$ and $R_2$ the respective rates of transmission.

Having access to perfect feedback the encoder can produce its time-$t$ channel inputs not only as a function of the messages $M_1$ and $M_2$ but also based on the previous channel outputs. Thus a block-length $n$ encoding scheme consists of $n$ functions $f_t^{(n)}$, for $t = 1, \ldots, n$, such that

$$X_t = f_t^{(n)}\left(M_1, M_2, \mathbf{Y}_1^{t-1}, \mathbf{Y}_2^{t-1}\right)$$

where $\mathbf{Y}_1^{t-1} \triangleq (Y_{1,1}, \ldots, Y_{1,t-1})$ and $\mathbf{Y}_2^{t-1} \triangleq (Y_{2,1}, \ldots, Y_{2,t-1})$. We impose an average block-power constraint $P > 0$ on the sequence of channel inputs:

$$\frac{1}{n} \mathsf{E}\left[\sum_{t=1}^{n} X_t^2\right] \leq P. \quad (3)$$

Based on the observed sequence of channel outputs $\mathbf{Y}_1^n$ and $\mathbf{Y}_2^n$, respectively, the two receivers perform the following guess of their corresponding message:

$$\hat{M}_k = \phi_k^{(n)}(\mathbf{Y}_k^n), \quad k \in \{1, 2\} \quad (4)$$

for some decoding functions $\phi_k^{(n)}$ for $k \in \{1, 2\}$.

An error occurs in the communication whenever $(M_1, M_2) \neq (\hat{M}_1, \hat{M}_2)$. We say that a rate pair $(R_1, R_2)$ is achievable if for every block-length $n$ there exist encoding functions $\{f_1^{(n)}, \ldots, f_n^{(n)}\}$ satisfying (3) and two decoding functions $\phi_1^{(n)}$ and $\phi_2^{(n)}$ such that

$$\lim_{n \to \infty} \Pr\left[(M_1, M_2) \neq (\hat{M}_1, \hat{M}_2)\right] = 0.$$

Of particular interest to this note is the *sum-rate capacity* $C(P, \sigma_1^2, \sigma_2^2, \rho_z)$, namely, the supremum of $R_1 + R_2$ for which reliable communication is feasible, i.e., where the pair $(R_1, R_2)$ is achievable.

## III. THE MAIN RESULT

The main result of this note concerns the so-called "pre-log", defined as follows.

*Definition 1:* Letting the sum-rate capacity be given by $C(P, \sigma_1^2, \sigma_2^2, \rho_z)$, its corresponding pre-log is defined as

$$\kappa = \varlimsup_{P \to \infty} \frac{C(P, \sigma_1^2, \sigma_2^2, \rho_z)}{\frac{1}{2} \log_2(1+P)}. \quad (5)$$

In the context of fading communication channels, the pre-log is often referred to as the *multiplexing gain.*

We start by noting that a pre-log of one is trivially attainable. Moreover, from the fact that for a broadcast channel without feedback, the capacity region only depends on the conditional marginals (see e.g. [11, p.599]), we have:

*Lemma 1:* For the two-user AWGN broadcast channel without feedback, the pre-log is 1 irrespective of the noise correlation $\rho_z$.

Also, by merely merging the two decoders into a single decoder, thus turning the problem into a point-to-point communication system, we find:

*Lemma 2:* For the two-user AWGN broadcast channel with full (causal) feedback, if the noise correlation satisfies $|\rho_z| < 1$, then the pre-log is 1.

The main result of this note is the following:

*Theorem 1:* For the two-user AWGN broadcast channel with full (causal) feedback, if the noise correlation is $\rho_z = -1$, then the pre-log is two.

The converse follows trivially by observing that with or without feedback, the following simple "single-user" upper bounds hold:

$$R_k \leq \frac{1}{2} \log_2\left(1 + \frac{P}{\sigma_1^2}\right), \quad k \in \{1, 2\}. \quad (6)$$

Thus, the pre-log cannot exceed two.

The somewhat more interesting part of the theorem concerns the achievability. The proof is given in Appendix A and is based on a strategy by Ozarow [5], [6] (see Section V).

## IV. SOME EXTENSIONS

### A. Limited Feedback

In the broadcast setting it can be shown that even if only one of the two channel outputs are fed back, a pre-log of two is attainable for the case of fully anti-correlated noises. This follows directly by noting that in this case one can compute one of the channel outputs based on the channel input and on the other channel output.

### B. More than Two Receivers

Consider a real scalar AWGN broadcast channel with $K > 2$ receivers. It can be shown that for more than two receivers Lemma 2 and Theorem 1 do not scale with the number of receivers, i.e., the maximum pre-log remains two irrespective of $K$. More precisely, let $\{Z_{k,t}\}$ denote the noise sequence corrupting the outputs of Receiver $k$, for $k \in \{1, \ldots, K\}$. Then, extending Lemma 2 and Theorem 1 to $K > 2$ receivers, the following two results can be derived: If for any $k, k' \in$ $\{1, \ldots, K\}$ with $k \neq k'$ the sequences $\{Z_{k,t}\}$ and $\{Z_{k',t}\}$ are not perfectly positively correlated or anti-correlated, then the pre-log equals one; if for any $k, k' \in \{1, \ldots, K\}$ with $k \neq k'$ the sequences $\{Z_{k,t}\}$ and $\{Z_{k',t}\}$ are not perfectly correlated and if additionally there exist $k_1, k_2 \in \{1, \ldots, K\}$ such that $\{Z_{k_1,t}\}$ and $\{Z_{k_2,t}\}$ are perfectly anti-correlated, then the pre-log equals two.

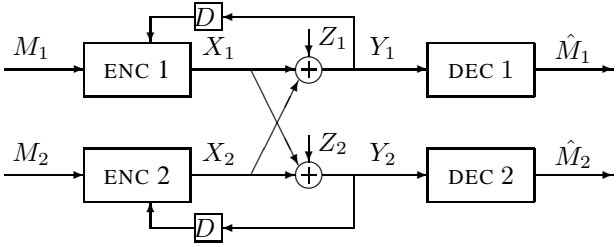

Fig. 2. The interference channel with ipsilateral causal output feedback.

### C. Interference Channel

An extension of our result concerns the two-user Gaussian *interference* channel, see Figure 2. The main difference to the previously considered broadcast setting is that here two transmitters wish to communicate.

The goal of the transmission is that Transmitter 1 conveys message $M_1$ to Receiver 1 and Transmitter 2 conveys Message $M_2$ to Receiver 2, where $M_1$ and $M_2$ are defined as before.

We assume that all links are of unit-gain (even though our results require only all equal gains), and hence the channel is described as follows. For given time-$t$ channel inputs $x_{1,t}$ at Transmitter 1 and $x_{2,t}$ at Transmitter 2 the channel outputs at the two receiving terminals are given by

$$Y_{k,t} = x_{1,t} + x_{2,t} + Z_{k,t}, \quad k \in \{1, 2\}, \quad (7)$$

where the noise sequences $\{(Z_{1,t}, Z_{2,t})\}$ are as in Section II.

Having access to full causal output feedback of their respective channel outputs, the two encoders can produce their time-$t$ channel inputs as

$$X_{k,t} = f_{k,t}^{(n)}(M_k, \mathbf{Y}_k^{t-1}), \quad k \in \{1, 2\}.$$

for some sequences of encoding functions $f_{1,t}^{(n)}$ and $f_{2,t}^{(n)}$, for $t = 1, \ldots, n$. As in the broadcast setting we impose an average block-power constraint $P > 0$ on the sequences of channel inputs:

$$\frac{1}{n}\mathsf{E}\left[\sum_{t=1}^{n} X_{k,t}^2\right] \leq P, \quad k \in \{1, 2\}. \quad (8)$$

The notion of decoding functions, probability of error, achievable rate pairs, and sum-rate capacity are in analogy to Section II.

The main result in this section is that Lemma 2 and Theorem 1 can be extended also to two-user Gaussian interference channels with all unit-gains.

*Lemma 3:* For the two-user AWGN interference channel with all unit channel gains and with full (causal) feedback to both transmitters, if the noise correlation satisfies $|\rho_z| < 1$, then the pre-log is 1.

*Theorem 2:* For the two-user AWGN interference channel with all unit channel gains and with full (causal) feedback to both transmitters, if the noise correlation is $\rho_z = -1$, then the pre-log is two.

The converse follows simply by Theorem 1 because letting the two transmitters cooperate can only increase capacity. The achievability is based on the following observations: In the broadcast strategy leading to Theorem 1 (see Section V) the single transmitter sends a weighted sum of the current estimation errors at the two receivers. In our interference setting due to the feedback links Transmitter 1 can compute the estimation error of Receiver 1 and Transmitter 2 can compute the estimation error of Receiver 2. Therefore, the two transmitters can mimic the single-transmitter strategy in Section V by each sending a scaled version of the corresponding estimation error because the channel implicitly adds up the two inputs. Hence, we can conclude that any rate pair achieved by the strategy in Section V for the broadcast channel is also achievable for the interference channel. Note however that it requires unit gain (or equal gain) on all channel links.

## V. Encoding Scheme and Analysis

The scheme we propose to prove the achievability of pre-log 2 in Theorem 1 follows along the lines of the scheme in [5], [6]. Our main contribution lies in the choice of the parameter $\gamma$ and the asymptotic analysis.

Just for completeness we give a short description of the scheme followed by a more detailed analysis of performance.

Prior to transmission, the encoder maps both messages with a one-to-one mapping into message points $\theta_1$ and $\theta_2$ in $(1/2, 1/2]$. More precisely,

$$\theta_\nu = 1/2 - \frac{M_\nu - 1}{\lfloor 2^{nR_\nu} \rfloor}, \quad \nu \in \{1, 2\}.$$

To start, the first channel use is dedicated to user 1 only, and the encoder transmits $\sqrt{\frac{P}{\mathsf{Var}(\theta_1)}}\theta_1$. The second channel use is dedicated to user 2 only, and the encoder transmits $\sqrt{\frac{P}{\mathsf{Var}(\theta_2)}}\theta_2$. Thereafter, each user forms an estimate of its message point, namely $\hat{\theta}_{1,1} = \sqrt{\frac{\mathsf{Var}(\theta_1)}{P}}Y_{1,1}$ and $\hat{\theta}_{2,2} = \sqrt{\frac{\mathsf{Var}(\theta_2)}{P}}Z_{2,2}$ respectively, incurring errors of

$$\epsilon_{1,2} = \sqrt{\frac{\mathsf{Var}(\theta_1)}{P}}Z_{1,1} \quad \text{and} \quad \epsilon_{2,2} = \sqrt{\frac{\mathsf{Var}(\theta_2)}{P}}Z_{2,2}. \quad (9)$$

In subsequent iterations, the encoder transmits a linear combination of the current receivers' estimation errors on $\theta_1$ and $\theta_2$, respectively. Thus, at time $k$ the channel input is

$$X_k = \sqrt{\frac{P}{1 + \gamma^2 + 2\gamma|\rho_{k-1}|}} \quad (10)$$

$$\cdot \left(\frac{\epsilon_{1,k-1}}{\sqrt{\alpha_{1,k-1}}} + \gamma\mathrm{sign}(\rho_{k-1})\frac{\epsilon_{2,k-1}}{\sqrt{\alpha_{2,k-1}}}\right) \quad (11)$$

where $\epsilon_{1,k-1}$ and $\epsilon_{2,k-1}$ denote the receiver's estimation error of $\theta_1$ and $\theta_2$ after the observation of the $(k-1)$-th channel output; where $\alpha_{1,k-1}$ and $\alpha_{2,k-1}$ denote the variances of the estimation errors and $\rho_{k-1}$ denotes their correlation coefficient; where sign($\cdot$) denotes the signum function, i.e., sign($x$) = 1 if $x \geq 0$ and sign($x$) = $-1$ otherwise; and where we choose (possibly sub-optimally)

$$\gamma = \frac{\sigma_1}{\sigma_2}. \quad (12)$$

After the reception of each $k$-th channel output each receiver performs a linear minimum mean square estimation (LMMSE) to estimate the respective error $\epsilon_{1,k-1}$ and $\epsilon_{2,k-1}$, and based on it they update their estimate of the respective message point.

At the end of each block of $n$ channel uses, each decoder guesses that the message has been transmitted which corresponds to the message point closest to its final estimate.

### A. Analysis

A detailed analysis of performance can be found in [5], [6]. Here we present the most important quantities of the analysis: the variances of the estimation errors at time-$k$

$$\alpha_{1,k} = \alpha_{1,k-1} \frac{\frac{\sigma_1^2}{\sigma_2^2} P(1-\rho_{k-1}^2) + (\sigma_1^2 + \sigma_2^2 + 2\sigma_1\sigma_2|\rho_{k-1}|)}{(1 + \frac{\sigma_1^2}{\sigma_2^2} + 2\frac{\sigma_1}{\sigma_2}|\rho_{k-1}|)(P + \sigma_1^2)}$$

and

$$\alpha_{2,k} = \alpha_{2,k-1} \frac{P(1-\rho_{k-1}^2) + (\sigma_1^2 + \sigma_2^2 + 2\sigma_1\sigma_2|\rho_{k-1}|)}{(1 + \frac{\sigma_1^2}{\sigma_2^2} + 2\frac{\sigma_1}{\sigma_2}|\rho_{k-1}|)(P + \sigma_2^2)},$$

and the correlation coefficient, see (13) on top of the next page. Note that Recursion (13) has at least one "fix point" $\rho^*$ in the interval $[0,1]$ in the sense that if $\rho_2 = \rho$ then the sequence $\{\rho_k\}$ alternates in sign but is constant in magnitude. This can seen by noting that for $\rho_{k-1} = 0$ it follows that $|\rho_k| > 0$ and for $|\rho_{k-1}| = 1$ it follows that $|\rho_k| < 1$, and thus by the continuity of the recursion there must exist a "fix point" $\rho \in [0,1]$.

By a slight modification of the scheme as suggested in [3] one can ensure that $\rho_2$ equals the "fix point" $\rho$ and one can show that any non-negative rate pair $(R_1, R_2)$ is achievable if it satisfies

$$R_1 < \frac{1}{2} \log_2 \left( \frac{P + \sigma_1^2}{\frac{P}{2}(1-\rho) + \sigma_1^2} \right) \quad (14)$$

$$R_2 < \frac{1}{2} \log_2 \left( \frac{P + \sigma_2^2}{\frac{P}{2}(1-\rho) + \sigma_2^2} \right) \quad (15)$$

where $\rho$ is a solution in $[0,1]$ of

$$\rho^3 + a\rho^2 + b\rho + c = 0 \quad (16)$$

where

$$a = -\frac{2\sigma_1\sigma_2}{P} - \frac{P + \sigma_1^2 + \sigma_2^2 + \rho_z\sigma_1\sigma_2}{\sqrt{P+\sigma_1^2}\sqrt{P+\sigma_2^2}}$$
$$- \frac{2\sigma_1^2\sigma_2^2}{P\sqrt{P+\sigma_1^2}\sqrt{P+\sigma_2^2}}, \quad (17)$$

$$b = -1 - \frac{\sigma_1^2 + \sigma_2^2}{P} - \rho_z \frac{(\sigma_1^2 + \sigma_2^2)}{\sqrt{P+\sigma_1^2}\sqrt{P+\sigma_2^2}}$$
$$- \frac{\sigma_1\sigma_2(\sigma_1^2 + \sigma_2^2)}{P\sqrt{P+\sigma_1^2}\sqrt{P+\sigma_2^2}}, \quad (18)$$

$$c = \frac{P + \sigma_1^2 + \sigma_2^2 - \rho_z\sigma_1\sigma_2}{\sqrt{P+\sigma_1^2}\sqrt{P+\sigma_2^2}}. \quad (19)$$

## APPENDIX

### A. Proof of Theorem 1

In this section we prove that for $\rho_z = -1$ the scheme described in Section V achieves a pre-log of 2.

The proof of the theorem follows directly by the achievability of rate pairs $(R_1, R_2)$ satisfying (14) and (15) and the following lemma.

*Lemma 4:* For $\rho_z = -1$ the function $\rho(P)$ implicitly defined by solutions in [0,1] to (16) satisfies

$$\lim_{P \to \infty} P^{1-\delta}(1-\rho(P)) = 0, \quad \forall \delta > 0. \quad (20)$$

*Proof:* Note first that the function $\rho(P)$ must satisfy

$$\lim_{P \to \infty} \rho(P) = 1. \quad (21)$$

This follows by the continuity of the coefficients $a$, $b$, and $c$ in $P$, and by observing that for large $P$ Equation (16) tends to $\rho^3 - \rho^2 - \rho + 1 = 0$ for which the only solutions are $-1$ and $+1$.

Next, define the function $g(P) \in [0,1]$ as

$$g(P) \triangleq 1 - \rho(P). \quad (22)$$

By (22) and (16) the function $g(P)$ must satisfy

$$0 = 1 + a + b + c + g(P)(-3 - 2a - b)$$
$$+ (g(P))^2 (3 + a) - (g(P))^3, \quad (23)$$

or equivalently by (17)–(19),

$$0 = -(g(P))^3 + \Lambda_2(P)(g(P))^2 + \Lambda_1(P)g(P) + \Lambda_0(P) \quad (24)$$

where

$$\Lambda_2(P) = 3 - \frac{2\sigma_1\sigma_2}{P} - \frac{P + \sigma_1^2 + \sigma_2^2 + \rho_z\sigma_1\sigma_2}{\sqrt{P+\sigma_1^2}\sqrt{P+\sigma_2^2}}$$
$$- \frac{2\sigma_1^2\sigma_2^2}{P\sqrt{P+\sigma_1^2}\sqrt{P+\sigma_2^2}},$$

$$\Lambda_1(P) = -2\left(1 - \frac{P}{\sqrt{P+\sigma_1^2}\sqrt{P+\sigma_2^2}}\right)$$
$$+ \frac{(2+\rho_z)\sigma_1^2 + (2+\rho_z)\sigma_2^2 + 2\rho_z\sigma_1\sigma_2}{\sqrt{P+\sigma_1^2}\sqrt{P+\sigma_2^2}}$$
$$+ \frac{\sigma_1^2 + \sigma_2^2 + 4\sigma_1\sigma_2}{P} + \frac{\sigma_1\sigma_2(\sigma_1^2 + 4\sigma_1\sigma_2 + \sigma_2^2)}{P\sqrt{P+\sigma_1^2}\sqrt{P+\sigma_2^2}},$$

$$\Lambda_0(P) = -\frac{\sigma_1^2 + 2\sigma_1\sigma_2 + \sigma_2^2}{P}\left(1 + \rho_z\frac{P}{\sqrt{P+\sigma_1^2}\sqrt{P+\sigma_2^2}}\right)$$
$$- \frac{\sigma_1\sigma_2(\sigma_1^2 + 2\sigma_1\sigma_2 + \sigma_2^2)}{P\sqrt{P+\sigma_1^2}\sqrt{P+\sigma_2^2}}.$$

$$\rho_k = \frac{\sqrt{P+\sigma_1^2}\sqrt{P+\sigma_2^2}}{P(1-\rho_{k-1}^2)+(\sigma_1^2+\sigma_2^2+2\sigma_1\sigma_2|\rho_{k-1}|)} \cdot \frac{1}{\sigma_1\sigma_2}$$
$$\cdot \left(\rho_{k-1}(\sigma_1^2+\sigma_2^2+2\sigma_1\sigma_2|\rho_{k-1}|) - (\sigma_1+\sigma_2|\rho_{k-1}|)(\sigma_2+\sigma_1|\rho_{k-1}|)\text{sign}(\rho_{k-1})\frac{P(P+\sigma_1^2+\sigma_2^2-\rho_z\sigma_1\sigma_2)}{(P+\sigma_1^2)(P+\sigma_2^2)}\right) \quad (13)$$

In the remaining we will prove that
$$\lim_{P\to\infty} P^{1-\delta}g(P) = 0, \quad \forall \delta > 0, \quad (25)$$
which by (22) establishes Lemma 4 and thus also concludes the proof of Theorem 1.

Start the proof by noting that
$$\varliminf_{P\to\infty} P^{1-\delta}g(P) \geq 0, \quad \forall \delta > 0,$$
follows trivially since $g(P) \geq 0$. Thus, we are left with proving
$$\varlimsup_{P\to\infty} P^{1-\delta}g(P) \leq 0, \quad \forall \delta > 0, \quad (26)$$
which we shall prove by contradiction. More precisely, we will show that if there exists a $\delta > 0$ such that $g(P)$ satisfies
$$\varlimsup_{P\to\infty} P^{1-\delta}g(P) > 0, \quad (27)$$
then Condition (24) on the function $g(P)$ is violated. To this end, assume that there exists a $\delta > 0$ satisfying (27). Then, define
$$\delta^* \triangleq \sup\left\{\delta : \varlimsup_{P\to\infty} P^{1-\delta}g(P) > 0\right\}, \quad (28)$$
and note that by assumption, $\delta^* > 0$. Also, by (21) and (22)
$$\lim_{P\to\infty} g(P) = 0, \quad (29)$$
and hence $\delta^* \leq 1$. Next, choose $0 < \epsilon < \delta^*$ and consider the asymptotic expression
$$\Delta \triangleq \varlimsup_{P\to\infty} P^{2-\delta^*-\epsilon}\left(-(g(P))^3 + \Lambda_2(P)(g(P))^2 + \Lambda_1(P)g(P) + \Lambda_0(P)\right). \quad (30)$$

We shall analyze the limiting expression (30) and show that it tends to a non-zero value. But this violates Condition (24) and therefore leads to the desired contradiction. In the analysis of expression (30) we shall separately consider the sum of the first two summands, the third summand, and forth summand. We start with the sum of the first two summands. Note first that
$$\lim_{P\to\infty} \Lambda_2(P) = 2. \quad (31)$$
Next, note that since $1 - \frac{\delta^*+\epsilon}{2} > 1 - \delta^*$ by (28)
$$\varlimsup_{P\to\infty} \left(P^{1-\frac{\delta^*+\epsilon}{2}}g(P)\right)^2 > 0, \quad (32)$$
Therefore, also using (29) we can conclude that
$$\varlimsup_{P\to\infty} \left(P^{1-\frac{\delta^*+\epsilon}{2}}g(P)\right)^2(\Lambda_2(P) - g(P)) > 0. \quad (33)$$

In order to analyze the third summand we notice that
$$\lim_{P\to\infty} P^{1-\epsilon/2}\Lambda_1(P) = 0, \quad (34)$$
where we used that by Bernoulli-de l'Hôpital's rule:
$$\lim_{P\to\infty} P\left(1 - \frac{P}{\sqrt{P+\sigma_1^2}\sqrt{P+\sigma_2^2}}\right) = \frac{\sigma_1^2+\sigma_2^2}{2}. \quad (35)$$
Hence, by (34) and because $1 - \delta^* - \epsilon/2 < 1 - \delta^*$,
$$\varlimsup_{P\to\infty} P^{2-\delta^*-\epsilon}\Lambda_1(P)g(P) = 0. \quad (36)$$
Finally, for the last summand one can show that for $\rho_z = -1$
$$\lim_{P\to\infty} P^{2-\delta^*-\epsilon}\Lambda_0(P) = 0, \quad (37)$$
which is again based on the limiting expression (35).

Thus, by (33), (36), and (37) we obtain that $\Delta > 0$ which contradicts Condition (24). This concludes the proof both of Lemma 4 and Theorem 1. ∎

ACKNOWLEDGMENT

The authors thank Prof. Frans M. J. Willems, TU Eindhoven, for pointing them to [7], which inspired the investigation leading to this work.